\documentclass[twocolumn,secnumarabic,amsmath, amssymb, nobibnotes, superscriptaddress, aps, prl]{revtex4-1}

\usepackage{graphicx}
\usepackage{verbatim}
\usepackage{hyperref}

\begin{document}

\title{Collective Suppression of Linewidths in Circuit QED}

\author{Felix Nissen}
\affiliation{Cavendish Laboratory, University of Cambridge, Cambridge CB3 0HE, United Kingdom}

\author{Johannes M.~Fink}
\altaffiliation[Now at ]{Institute for Quantum Information and Matter, California Institute of Technology, Pasadena, CA 91125, USA}
\affiliation{Department of Physics, ETH Zurich, CH-8093 Zurich, Switzerland}

\author{Jonas A.~Mlynek}
\affiliation{Department of Physics, ETH Zurich, CH-8093 Zurich, Switzerland}

\author{Andreas Wallraff}
\affiliation{Department of Physics, ETH Zurich, CH-8093 Zurich, Switzerland}

\author{Jonathan Keeling}
\affiliation{Scottish Universities Physics Alliance, School of Physics and Astronomy, University of St Andrews, St Andrews KY16 9SS, United Kingdom}

\begin{abstract}
  We report the experimental observation, and a theoretical explanation,
  of collective suppression of linewidths for multiple superconducting
  qubits coupled to a good cavity.  This demonstrates how strong
  qubit-cavity coupling can significantly modify the dephasing and
  dissipation processes that might be expected for individual qubits,
  and can potentially improve coherence times in many-body circuit
  QED.
\end{abstract}
\maketitle

At the root of many of the unexpected effects predicted by quantum
mechanics is quantum interference.  In the context of quantum optics,
one notable consequence of constructive interference is
superradiance~\cite{Dicke1954,Gross1982}, the collective enhancement
of radiation from an ensemble of many initially excited atoms.  At its
heart is the idea that if the atoms remain in a symmetric state, then
constructive interference between different final states after
emitting a photon can increase the probability of such photon emission
events~\cite{Dicke1954}.  It is notable that such constructive
interference plays a role even when considering the incoherent and
irreversible emission of photons into an external environment. This collective effect is distinct from the subnatural linewidth averaging first discussed by \cite{carmichael1989subnatural}.

Since the superradiant emission of photons can occur incoherently, it
can be expected that similar effects should be visible in the
linewidth of a collection of atoms, or artificial atoms, coupled
symmetrically to the same solid state environment.  Indeed, one may see such an
effect by comparing collective and individual decay processes.
Consider $N$ two-level systems (2LS) obeying either individual decay
and dephasing: $\dot{\rho} = \dot{\rho}_H + \sum_i
(\gamma_{\parallel}/2)\mathcal{D}[\sigma_i^z] +
(\gamma_{\perp}/2)\mathcal{D}[\sigma_i^-]$ (where $\mathcal{D}[X] = 2
X \rho X^\dagger - X^\dagger X \rho - \rho X^\dagger X$, $\{\sigma_i\}$ are Pauli operators describing the two-level systems with associated dephasing ($\gamma_{\parallel}$) and relaxation ($\gamma_{\perp}$) rates, and
$\dot{\rho}_H = -i[H,\rho]$ describes the Hamiltonian evolution), or
collective decay and dephasing: $\dot{\rho} = \dot{\rho}_H +
(\gamma_{\parallel}/2)\mathcal{D}[\sum_i \sigma_i^z] +
(\gamma_{\perp}/2)\mathcal{D}[\sum_i \sigma_i^-]$.  If one then
calculates the linear absorption spectrum for an environment coupling
symmetrically to all 2LS one finds a total linewidth $1/T_2 = 2
\gamma_{\parallel} + \gamma_{\perp}/2$ for the case of individual
decay, and $1/T_2 = 2 \gamma_{\parallel} + N \gamma_{\perp}/2$ for collective decay.
The linewidth associated with the coupling to a common bath is
collectively enhanced, because of the constructive interference of
different decay pathways.

A natural context in which such questions arise is solid state
realizations of coupled matter-light systems, such as
circuit-QED~\cite{PhysRevA.69.062320,wallraff2004strong}, where multiple superconducting qubits
can be confined in a single microwave
cavity~\cite{Wallraff:Collective,Majer2007,DiCarlo:Prep}, and so may
potentially couple to a common reservoir.  In the limit of a good
cavity, where a significant part of the vacuum Rabi linewidths is due to non
cavity-mediated decay and dissipation, the distinction between
coupling to collective and separate decay channels should be apparent
in the dependence of linewidth on the number of qubits present.  Even
within a single sample, the effective qubit number can be easily
varied by detuning the qubits away from resonance with each other~\cite{Wallraff:Collective}.
This breaks the symmetry, providing which-path information, and
thus destroys the coherence.  The question of whether decay and dephasing
of multiple qubits is due to separate or collective coupling to the
environment may have important consequences for the ability to
preserve and manipulate coherence.  For example, unexpectedly long
coherence in light-harvesting-complexes~\cite{Collini2010} is
associated with non-trivial quantum dynamics arising from coupling to
common photon modes~\cite{Semiao2010,*Chin}.

There is however a problem with the simple picture of collective
enhancement of linewidth when applied to multiple qubits coupled to a
microwave cavity.  The problem is that the Lindblad terms written in
the above are those that would be derived by considering
system-reservoir coupling where the system Hamiltonian is that of a
single qubit.  The importance of using the correct system Hamiltonian
in deriving loss terms has long been recognised in the context of
ensuring that the correct equilibrium state is reached
asymptotically~\cite{Carmichael1973,Cresser1992}.  More recently there
has been significant activity on such issues in the context of quantum
dots
\cite{Wilson-Rae2002,Forstner2003,Vagov2007a,Ramsay2010a,Ramsay2010,McCutcheon2010a,Roy2011,Luker2012,Eastham2012,Reiter2012}.
In particular, it has been noted that when the system Hamiltonian is
significantly changed, either by external driving
\cite{Forstner2003,Vagov2007a,Ramsay2010a,Ramsay2010,McCutcheon2010a,Roy2011,Luker2012,Eastham2012,Reiter2012},
or (as in the current case) strong qubit-cavity
coupling\cite{Wilson-Rae2002,Roy2011}, it is crucial to recalculate
the decay processes in the presence of the \emph{full} system
Hamiltonian.  This is particularly important when the reservoir has a
non-flat frequency dependence.  While a sufficiently rapidly varying
frequency dependence may prevent a Markovian density matrix being used
at all, there is a significant range of parameters where a Born-Markov
approach remains valid, but it is necessary to calculate the decay
rates using the correct system Hamiltonian, rather than regarding the
decay rates as fixed parameters.  It is clear that the system of many
qubits coupled to a common photon mode is such a case: When the
coupling and number of systems becomes large enough, the system has
been predicted ~\cite{Nataf:Nogo} to undergo a
phase transition to a spontaneously polarized state~\footnote{There is some debate
  regarding whether the transition can be achieved for circuit
  QED~\cite{Rzazewski1975a,*Viehmann:SR,*Vukics2012}.}; at this point the
frequency of the collective mode should vanish, and
care~\cite{Ciuti2006} must be taken to avoid unphysical predictions. The importance of using the eigenstates of strongly-coupled Hamiltonians has also been pointed out in the context of quantising superconducting circuits \cite{PhysRevLett.108.240502}.

In this letter we report measurements and calculations of the
linewidth for one, two and three qubits coupled resonantly to a
microwave cavity. In contrast to the well studied effect of superradiance we observe and explain the \emph{narrowing of linewidths}
in the strong dephasing regime (i.e. dephasing and cavity decay rates are comparable) as the number of
qubits is increased.  We find the results are compatible
with a model of ``collective'' dephasing processes, where all qubits
couple to a single bath, with a spectrum corresponding to $1/f$
noise.  We also discuss how varying the cavity-qubit detuning might
allow corroboration of this scenario, and a direct measurement of
the dephasing bath spectrum.

\begin{figure}[t]
\centering
\includegraphics[height=3.2in,angle=-90]{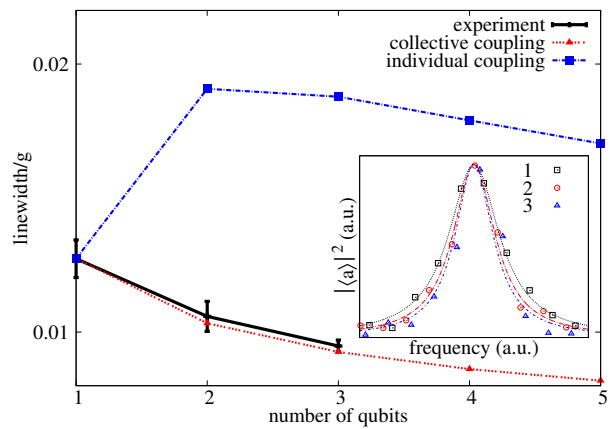}
\caption{Experimental linewidth (black, solid) and theoretical
  linewidths (red and blue dashed) of an $N$-qubit-cavity system
  (error bars indicate the sample standard deviation of the fitted
  linewidths). Theoretical linewidths are calculated for either collective
  coupling (lower line) or individual
  coupling (upper line) to a reservoir with a $1/f$ spectrum (density of states $J(\nu>0)=0.0105g/\nu$). The inset shows the experimental spectrum of the coherently
  scattered transmission amplitude for $1$, $2$ and $3$ qubits, re-scaled and shifted so that
  peaks are centered at the same frequency for
  comparison. The lines are the corresponding Lorentzian fits. For every number $N$ of qubits, both Rabi peaks were measured and used to calcuate a single linewidth and its uncertainty. The inset only shows one set of Rabi peaks.
  \label{fig:lw}}
\end{figure}

Figure~\ref{fig:lw} shows part of the vacuum Rabi transmission spectrum (inset) and
the extracted linewidths measured for a microwave cavity with one, two and
three qubits tuned into resonance with the cavity mode. The superconducting
microchip sample and setup is similar to the one used in Refs.~\cite{Wallraff:Collective,Mlynek2012}.
The transmission spectrum is measured with much less than a single intra-cavity photon on average and clearly shows the expected 
$\sqrt{N}$ dependence of the vacuum Rabi splitting with increasing number of qubits (not
shown). The single qubit-photon couplings of the three superconducting transmon qubits are $g/(2\pi)=(52.7,55.4,55.8)$ MHz. The over-coupled coplanar waveguide resonator has a first harmonic resonance frequency of $\nu_r = 7.0235$ GHz and a quality factor of $Q = 14800$ as measured when the qubits are far detuned (corresponding to a cavity decay rate $\kappa/2\pi=0.47\,\mathrm{MHz}$). Time-resolved off-resonant $T_1$ and $T_2$ measurements of the three qubits confirm that $T_1\gg T_2\approx 150\,\mathrm{ns}$ and that the qubit linewidths have strong dephasing components, i.e. $1/T_2\approx\kappa$ for each qubit is fulfilled. The dressed linewidths on resonance are therefore expected to be dominated by qubit dephasing as well.
While there is a considerable experimental uncertainty to the individual measurement data, see inset in Fig.~\ref{fig:lw}, the extracted linewidth from 6 single qubit, 2 two qubit and 2 three qubit Rabi peaks (as indicated with error bars), shows a very clear trend of linewidth narrowing as the number of resonant qubits is increased.

As discussed above, if the dephasing could be modeled by Lindblad
operators derived for the uncoupled system, then narrowing of the
linewidth would be unexpected. For a fixed dephasing rate, one would expect
a constant or increasing linewidth.  As discussed below, one can
however directly explain this narrowing as a result of coupling the
full system Hamiltonian to a frequency dependent bath; we assume in the following that dephasing is due to a bath with a $1/f$ spectrum~\cite{Yoshihara2006a}. The
frequency at which the coupled system samples the bath depends on the
energy differences between system eigenstates. These depend on the
collective Rabi frequency, which scales as $\sqrt{N}$ for $N$ qubits.  Thus, the
effective decay rate decreases with increasing number of qubits.  This
simple argument explains the essential origin of the results seen
experimentally, however several complications occur when one actually
calculates the effective linewidth using the strongly-coupled
qubit-resonator Hamiltonian.

As discussed also above, two different scenarios of dephasing can
exist; individual coupling to reservoirs and collective coupling.
When accounting for the qubit-resonator coupling, the dephasing
induced linewidth depends differently on $N$ in these two cases.  In
the collective coupling case, the linewidth scales as $1/\sqrt{N}$ as
the simple argument given above suggests: The dominant effect is the
sampling the reservoir at the collective Rabi frequency.  Small
corrections occur due to the matrix elements between system
eigenstates arising from coupling to the bath and the small but non-zero photon decay rates.  In the case of
individual coupling, $1/\sqrt{N}$ is still the dominant effect, but as
system eigenstates are now delocalized, a competing effect arises: The
number of possible decay channels increases with qubit number, as
cross-qubit terms induced by the resonator couple qubits to other
baths. This gives rise to an initial increase, followed by a decrease
of the linewidth.  Figure~\ref{fig:lw} shows the results of the
linewidth vs number of qubits in these two cases of individual and
collective decay, assuming a $1/f$ spectrum for the reservoir.
Further details of the calculation are given in the following.

\begin{figure}[tpb]
\centering
\includegraphics[height=3.2in,angle=-90]{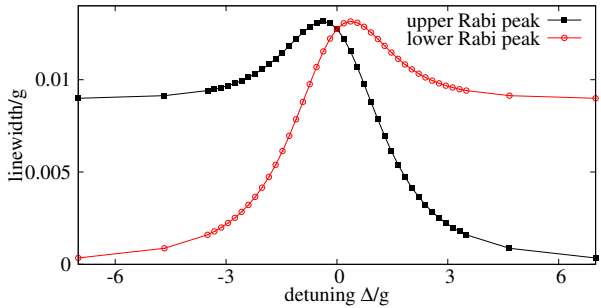}
\caption{Relative theoretical linewidth of a single qubit-cavity
  system as a function of resonator-qubit detuning. The linewidths of
  both Rabi peaks (squares: upper peak, circles: lower peak) decrease because the photon part has a small decay
  rate and the qubit part samples the $1/f$ decay bath at frequencies
  $\sqrt{\Delta^2 + g^2}$. \label{fig:theo-detuning}}
\end{figure}

In order to disentangle the effects of the reservoir density of states
from the effects of the cross-coupling matrix elements, one may
instead explore the dependence of linewidth on cavity-qubit detuning.
For a single qubit coupled strongly to a cavity, the normal modes are
superpositions of photon and qubit states.  As the qubit-cavity
detuning $\Delta$ is varied two effects occur: The nature of the modes
changes, and the frequency at which the reservoir is sampled varies as the Rabi
frequency $\sqrt{\Delta^2+g^2}$.  The changing nature of the modes
means that one mode becomes more photon-like, and the other will have
a large qubit weight.  The changing Rabi frequency means that the
reservoir is sampled at higher frequencies, and so for a $1/f$
spectrum, the linewidth will decrease.  The calculated linewidth vs
detuning for a single qubit is shown in Fig.~\ref{fig:theo-detuning}.
Both the changing nature of the modes (causing the linewidths of the
two modes to differ) and the reduction of the qubit dephasing occur on
a similar scale, $\Delta \sim g$, as is clear from the figure. For
multiple qubits, the detuning dependence of the linewidths still
depends on the nature of the coupling (individual vs collective) as
well as the density of states of the reservoir.  Nonetheless, the
detuning dependence of linewidth can corroborate a given model of
dephasing, and provide clear information about the real spectrum of
the environment which induces dephasing.

We now turn to discuss in more detail how the dephasing rate is
calculated using the Born-Markov approximation~\cite{Breuer2007},
with the full Hamiltonian.  Using Pauli operators
$\sigma_i^{x,y,z}$ to represent the qubits and bosonic operators
$a,a^\dagger$ for the cavity, we have
\begin{align}
H_\mathrm{sys} &= \omega a^\dagger a + \sum_{i=1}^3 \frac{\epsilon_i}{2} \sigma_i^z + \sum_{i=1}^3 g \left(\sigma_i^+ a + \mathrm{h.c.}\right)\\
H_\mathrm{bath} &= \sum_{i,q}
\gamma_q\sigma_i^z\left(b_{iq}^\dagger+b_{iq}\right)
+ \beta_{iq} b_{iq}^\dagger b_{iq}.
\end{align}
$b_{iq}$ are the bosonic modes of the environment with energy $\beta_{iq}$ whose coupling strength to the system is $\gamma_{q}$. The important quantity for the behaviour of the qubits is the combination $J_i(\nu) = \sum_q\gamma_{iq}^2\delta(\nu-\beta_{iq})$. The total Hamiltonian $H= H_\mathrm{sys} + H_\mathrm{bath}$ describes
the strongly coupled qubit-resonator system and the dephasing term in
the case of coupling to individual reservoirs.  For collective
coupling one instead has $H_\mathrm{bath} =
\left(\sum_{i}\sigma_i^z\right)\sum_{q}\gamma_q\left(b_{q}^\dagger+b_{q}\right)
+ \sum_{q} \beta_{q} b_{q}^\dagger b_{q}$.  In the interaction picture
the equation of motion is
\begin{multline}
  \dot{\rho} =
  \sum\limits_{i} \!
  \int\limits_{-\infty}^\infty\!\!\mathrm{d}\nu J(\nu)\!\!\!
  \int\limits_{-\infty}^t\!\!\mathrm{d}t'
  \biggl\{
  \left[ P_i(t')\rho P_i(t) - P_i(t) P_i(t')\rho\right]\times
  \biggr.\\\biggl.
   f_\nu(t-t')
  +
  \left[ P_i(t)\rho P_i(t') - \rho P_i(t')P_i(t)  \right]f_\nu(t'-t)\!
  \biggr\} \label{eq:eom}
\end{multline}
with $P_i(t) = e^{i H_\mathrm{sys} t} \sigma_i^z e^{-i H_\mathrm{sys}
  t}$ for individual reservoirs and $P(t) = e^{i H_\mathrm{sys} t}
\left(\sum_i \sigma_i^z\right) e^{-i H_\mathrm{sys} t}$ for collective
dephasing.  We assume the reservoir density of states to be
$J(\nu>0)\propto 1/\nu$ and $f_\nu(\tau) =f_\nu^*(-\tau) = (n_\nu+1) e^{-i
  \nu \tau} + n_\nu e^{i \nu \tau}$ depends on the Bose-Einstein
occupation $n_\nu$ which we may assume to be zero, since $k_B T \ll
g$.

By assuming $\rho=\rho(t)$ on the right hand side of
Eq. (\ref{eq:eom}) we make the standard Markov
approximation\cite{Breuer2007}, but retain the crucial dependence on
the collective Rabi frequency by evaluating the remaining integrals with
the full system Hamiltonian~\cite{Ramsay2010}.  For a general system
Hamiltonian, Eq.~(\ref{eq:eom}) is not in Lindblad form and so does
not necessarily preserve positivity of the density matrix
equation~\cite{Breuer2007}.  For short times, this is not an
issue~\cite{Ramsay2010}, but for long times (as matters for steady
states), positivity violation becomes a
problem~\cite{Eastham2012,Gawarecki2012}. One may however obtain a
Lindblad form by dropping any terms which are time-dependent in the
interaction picture~\cite{Dmcke1979}; the resulting time evolution is
equivalent to Eq.~(\ref{eq:eom}) over short times~\cite{Dmcke1979}, but
avoids unphysical positivity violations.  The resulting Lindblad terms
have the form:
\begin{equation}
  \mathcal{L}_\mathrm{d} =
  \sum\limits_{i\alpha\beta\gamma\delta}
  \left(
    2 r^i_{\gamma\delta}\rho r^{i\dagger}_{\beta\alpha} -
    r^{i\dagger}_{\beta\alpha} r^i_{\gamma\delta}\rho  -
    \rho r^{i\dagger}_{\beta\alpha} r^i_{\gamma\delta}
  \right) \label{eq:decay}
\end{equation}
with
$r_{\alpha\beta}^i=\sqrt{A_{\alpha\beta}}|\alpha\rangle\langle\alpha|\sigma_i^z|\beta\rangle\langle\beta|$
where $\alpha, \beta$ are eigenstates of the system Hamiltonian and
$A_{\alpha\beta} = \Gamma(\epsilon_\alpha-\epsilon_\beta)$ is a
transition-energy dependent decay rate, $\Gamma(\delta) = \pi \left[
  J(\delta)(n(\delta)+1) + J(-\delta) n(-\delta)\right]$.  The
summation over states is restricted to energy-conserving transitions
$E_\alpha-E_\beta=E_\delta-E_\gamma$.  For collective decay, the sum
over $i$ disappears, and $r_{\alpha\beta}$ is defined in terms of
matrix elements of the operator $\sum_i \sigma_i^z$ instead.

This dephasing term is of Lindblad form as required to preserve
positivity of the density matrix. However, it is not simply a collection
of qubit dephasing terms added \emph{ad hoc} to the density matrix
equation of motion. Instead, we naturally obtain dephasing in terms of
system eigenstates, sampling the reservoir density of states at the
system's natural transition frequencies.  To calculate the coherent
photon scattering spectrum we numerically solve the steady state of
this density matrix equation in the presence of a weak drive.  (Since
the drive is weak, it does not itself modify the decay rates).  The
spectrum is then found by calculating $|\langle a \rangle|$ vs
detuning, and the linewidth plotted in
Figs.~\ref{fig:lw} and \ref{fig:theo-detuning} is extracted by
fitting this spectrum to a Lorentzian.

In conclusion, we have shown how strong matter-light coupling and a
non-trivial reservoir spectrum can produce a non-trivial
\emph{suppression} of linewidth, which can nonetheless be explained
within a Markovian approximation.  Our calculations explain why
linewidths in an $N$-qubit-cavity system can decrease with the number
of qubits. We further make testable predictions for an off-resonant
qubit-cavity system and offer a way to probe the reservoir density of
states.

\acknowledgements{JK and FN acknowledge discussions with P. Eastham
  and A. Chin.  JK acknowledges financial support from EPSRC program
  ``TOPNES'' (EP/I031014/1) and EPSRC (EP/G004714/2).}


\begin{thebibliography}{35}%
\makeatletter
\providecommand \@ifxundefined [1]{%
 \@ifx{#1\undefined}
}%
\providecommand \@ifnum [1]{%
 \ifnum #1\expandafter \@firstoftwo
 \else \expandafter \@secondoftwo
 \fi
}%
\providecommand \@ifx [1]{%
 \ifx #1\expandafter \@firstoftwo
 \else \expandafter \@secondoftwo
 \fi
}%
\providecommand \natexlab [1]{#1}%
\providecommand \enquote  [1]{``#1''}%
\providecommand \bibnamefont  [1]{#1}%
\providecommand \bibfnamefont [1]{#1}%
\providecommand \citenamefont [1]{#1}%
\providecommand \href@noop [0]{\@secondoftwo}%
\providecommand \href [0]{\begingroup \@sanitize@url \@href}%
\providecommand \@href[1]{\@@startlink{#1}\@@href}%
\providecommand \@@href[1]{\endgroup#1\@@endlink}%
\providecommand \@sanitize@url [0]{\catcode `\\12\catcode `\$12\catcode
  `\&12\catcode `\#12\catcode `\^12\catcode `\_12\catcode `\%12\relax}%
\providecommand \@@startlink[1]{}%
\providecommand \@@endlink[0]{}%
\providecommand \url  [0]{\begingroup\@sanitize@url \@url }%
\providecommand \@url [1]{\endgroup\@href {#1}{\urlprefix }}%
\providecommand \urlprefix  [0]{URL }%
\providecommand \Eprint [0]{\href }%
\@ifxundefined \urlstyle {%
  \providecommand \doi  [0]{\begingroup \@sanitize@url \@doi}%
  \providecommand \@doi [1]{\endgroup \@@startlink {\doibase
  #1}doi:\discretionary {}{}{}#1\@@endlink }%
}{%
  \providecommand \doi  [0]{doi:\discretionary{}{}{}\begingroup
  \urlstyle{rm}\Url }%
}%
\providecommand \doibase [0]{http://dx.doi.org/}%
\providecommand \Doi [0]{\begingroup \@sanitize@url \@Doi }%
\providecommand \@Doi  [1]{\endgroup\@@startlink{\doibase#1}\@@Doi}%
\providecommand \@@Doi [1]{#1\@@endlink}%
\providecommand \selectlanguage [0]{\@gobble}%
\providecommand \bibinfo  [0]{\@secondoftwo}%
\providecommand \bibfield  [0]{\@secondoftwo}%
\providecommand \translation [1]{[#1]}%
\providecommand \BibitemOpen [0]{}%
\providecommand \bibitemStop [0]{}%
\providecommand \bibitemNoStop [0]{.\EOS\space}%
\providecommand \EOS [0]{\spacefactor3000\relax}%
\providecommand \BibitemShut  [1]{\csname bibitem#1\endcsname}%
\bibitem [{\citenamefont {Dicke}(1954)}]{Dicke1954}%
  \BibitemOpen
  \bibfield  {author} {\bibinfo {author} {\bibfnamefont {R.}~\bibnamefont
  {Dicke}},\ }\Doi {10.1103/PhysRev.93.99} {\bibfield  {journal} {\bibinfo
  {journal} {Phys. Rev.},\ }\textbf {\bibinfo {volume} {93}},\ \bibinfo {pages}
  {99} (\bibinfo {year} {1954})},\ ISSN \bibinfo {issn} {0031-899X}\BibitemShut
  {NoStop}%
\bibitem [{\citenamefont {Gross}\ and\ \citenamefont
  {Haroche}(1982)}]{Gross1982}%
  \BibitemOpen
  \bibfield  {author} {\bibinfo {author} {\bibfnamefont {M.}~\bibnamefont
  {Gross}}\ and\ \bibinfo {author} {\bibfnamefont {S.}~\bibnamefont
  {Haroche}},\ }\Doi {10.1016/0370-1573(82)90102-8} {\bibfield  {journal}
  {\bibinfo  {journal} {Phys. Rep.},\ }\textbf {\bibinfo {volume} {93}},\
  \bibinfo {pages} {301} (\bibinfo {year} {1982})},\ ISSN \bibinfo {issn}
  {03701573}\BibitemShut {NoStop}%
\bibitem [{\citenamefont {Carmichael}\ \emph {et~al.}(1989)\citenamefont
  {Carmichael}, \citenamefont {Brecha}, \citenamefont {Raizen}, \citenamefont
  {Kimble},\ and\ \citenamefont {Rice}}]{carmichael1989subnatural}%
  \BibitemOpen
  \bibfield  {author} {\bibinfo {author} {\bibfnamefont {H.}~\bibnamefont
  {Carmichael}}, \bibinfo {author} {\bibfnamefont {R.}~\bibnamefont {Brecha}},
  \bibinfo {author} {\bibfnamefont {M.}~\bibnamefont {Raizen}}, \bibinfo
  {author} {\bibfnamefont {H.}~\bibnamefont {Kimble}}, \ and\ \bibinfo {author}
  {\bibfnamefont {P.}~\bibnamefont {Rice}},\ }\href@noop {} {\bibfield
  {journal} {\bibinfo  {journal} {Physical Review A},\ }\textbf {\bibinfo
  {volume} {40}},\ \bibinfo {pages} {5516} (\bibinfo {year}
  {1989})}\BibitemShut {NoStop}%
\bibitem [{\citenamefont {Blais}\ \emph {et~al.}(2004)\citenamefont {Blais},
  \citenamefont {Huang}, \citenamefont {Wallraff}, \citenamefont {Girvin},\
  and\ \citenamefont {Schoelkopf}}]{PhysRevA.69.062320}%
  \BibitemOpen
  \bibfield  {author} {\bibinfo {author} {\bibfnamefont {A.}~\bibnamefont
  {Blais}}, \bibinfo {author} {\bibfnamefont {R.-S.}\ \bibnamefont {Huang}},
  \bibinfo {author} {\bibfnamefont {A.}~\bibnamefont {Wallraff}}, \bibinfo
  {author} {\bibfnamefont {S.~M.}\ \bibnamefont {Girvin}}, \ and\ \bibinfo
  {author} {\bibfnamefont {R.~J.}\ \bibnamefont {Schoelkopf}},\ }\Doi
  {10.1103/PhysRevA.69.062320} {\bibfield  {journal} {\bibinfo  {journal}
  {Phys. Rev. A},\ }\textbf {\bibinfo {volume} {69}},\ \bibinfo {pages}
  {062320} (\bibinfo {year} {2004})}\BibitemShut {NoStop}%
\bibitem [{\citenamefont {Wallraff}\ \emph {et~al.}(2004)\citenamefont
  {Wallraff}, \citenamefont {Schuster}, \citenamefont {Blais}, \citenamefont
  {Frunzio}, \citenamefont {Huang}, \citenamefont {Majer}, \citenamefont
  {Kumar}, \citenamefont {Girvin},\ and\ \citenamefont
  {Schoelkopf}}]{wallraff2004strong}%
  \BibitemOpen
  \bibfield  {author} {\bibinfo {author} {\bibfnamefont {A.}~\bibnamefont
  {Wallraff}}, \bibinfo {author} {\bibfnamefont {D.}~\bibnamefont {Schuster}},
  \bibinfo {author} {\bibfnamefont {A.}~\bibnamefont {Blais}}, \bibinfo
  {author} {\bibfnamefont {L.}~\bibnamefont {Frunzio}}, \bibinfo {author}
  {\bibfnamefont {R.}~\bibnamefont {Huang}}, \bibinfo {author} {\bibfnamefont
  {J.}~\bibnamefont {Majer}}, \bibinfo {author} {\bibfnamefont
  {S.}~\bibnamefont {Kumar}}, \bibinfo {author} {\bibfnamefont
  {S.}~\bibnamefont {Girvin}}, \ and\ \bibinfo {author} {\bibfnamefont
  {R.}~\bibnamefont {Schoelkopf}},\ }\href@noop {} {\bibfield  {journal}
  {\bibinfo  {journal} {Nature},\ }\textbf {\bibinfo {volume} {431}},\ \bibinfo
  {pages} {162} (\bibinfo {year} {2004})}\BibitemShut {NoStop}%
\bibitem [{\citenamefont {Fink}\ \emph {et~al.}(2009)\citenamefont {Fink},
  \citenamefont {Bianchetti}, \citenamefont {Baur}, \citenamefont {G\"{o}ppl},
  \citenamefont {Steffen}, \citenamefont {Filipp}, \citenamefont {Leek},
  \citenamefont {Blais}, \citenamefont {Wallraff},\ and\ \citenamefont
  {Go}}]{Wallraff:Collective}%
  \BibitemOpen
  \bibfield  {author} {\bibinfo {author} {\bibfnamefont {J.~M.}\ \bibnamefont
  {Fink}}, \bibinfo {author} {\bibfnamefont {R.}~\bibnamefont {Bianchetti}},
  \bibinfo {author} {\bibfnamefont {M.}~\bibnamefont {Baur}}, \bibinfo {author}
  {\bibfnamefont {M.}~\bibnamefont {G\"{o}ppl}}, \bibinfo {author}
  {\bibfnamefont {L.}~\bibnamefont {Steffen}}, \bibinfo {author} {\bibfnamefont
  {S.}~\bibnamefont {Filipp}}, \bibinfo {author} {\bibfnamefont {P.~J.}\
  \bibnamefont {Leek}}, \bibinfo {author} {\bibfnamefont {A.}~\bibnamefont
  {Blais}}, \bibinfo {author} {\bibfnamefont {A.}~\bibnamefont {Wallraff}}, \
  and\ \bibinfo {author} {\bibfnamefont {M.}~\bibnamefont {Go}},\ }\Doi
  {10.1103/PhysRevLett.103.083601} {\bibfield  {journal} {\bibinfo  {journal}
  {Phys. Rev. Lett.},\ }\textbf {\bibinfo {volume} {103}},\ \bibinfo {pages}
  {83601} (\bibinfo {year} {2009})}\BibitemShut {NoStop}%
\bibitem [{\citenamefont {Majer}\ \emph {et~al.}(2007)\citenamefont {Majer},
  \citenamefont {Chow}, \citenamefont {Gambetta}, \citenamefont {Koch},
  \citenamefont {Johnson}, \citenamefont {Schreier}, \citenamefont {Frunzio},
  \citenamefont {Schuster}, \citenamefont {Houck}, \citenamefont {Wallraff},
  \citenamefont {Blais}, \citenamefont {Devoret}, \citenamefont {Girvin},\ and\
  \citenamefont {Schoelkopf}}]{Majer2007}%
  \BibitemOpen
  \bibfield  {author} {\bibinfo {author} {\bibfnamefont {J.}~\bibnamefont
  {Majer}}, \bibinfo {author} {\bibfnamefont {J.~M.}\ \bibnamefont {Chow}},
  \bibinfo {author} {\bibfnamefont {J.~M.}\ \bibnamefont {Gambetta}}, \bibinfo
  {author} {\bibfnamefont {J.}~\bibnamefont {Koch}}, \bibinfo {author}
  {\bibfnamefont {B.~R.}\ \bibnamefont {Johnson}}, \bibinfo {author}
  {\bibfnamefont {J.~A.}\ \bibnamefont {Schreier}}, \bibinfo {author}
  {\bibfnamefont {L.}~\bibnamefont {Frunzio}}, \bibinfo {author} {\bibfnamefont
  {D.~I.}\ \bibnamefont {Schuster}}, \bibinfo {author} {\bibfnamefont {A.~A.}\
  \bibnamefont {Houck}}, \bibinfo {author} {\bibfnamefont {A.}~\bibnamefont
  {Wallraff}}, \bibinfo {author} {\bibfnamefont {A.}~\bibnamefont {Blais}},
  \bibinfo {author} {\bibfnamefont {M.~H.}\ \bibnamefont {Devoret}}, \bibinfo
  {author} {\bibfnamefont {S.~M.}\ \bibnamefont {Girvin}}, \ and\ \bibinfo
  {author} {\bibfnamefont {R.~J.}\ \bibnamefont {Schoelkopf}},\ }\Doi
  {10.1038/nature06184} {\bibfield  {journal} {\bibinfo  {journal} {Nature},\
  }\textbf {\bibinfo {volume} {449}},\ \bibinfo {pages} {443} (\bibinfo {year}
  {2007})},\ ISSN \bibinfo {issn} {1476-4687}\BibitemShut {NoStop}%
\bibitem [{\citenamefont {Dicarlo}\ \emph {et~al.}(2010)\citenamefont
  {Dicarlo}, \citenamefont {Reed}, \citenamefont {Sun}, \citenamefont
  {Johnson}, \citenamefont {Chow}, \citenamefont {Gambetta}, \citenamefont
  {Frunzio}, \citenamefont {Girvin}, \citenamefont {Devoret},\ and\
  \citenamefont {Schoelkopf}}]{DiCarlo:Prep}%
  \BibitemOpen
  \bibfield  {author} {\bibinfo {author} {\bibfnamefont {L.}~\bibnamefont
  {Dicarlo}}, \bibinfo {author} {\bibfnamefont {M.~D.}\ \bibnamefont {Reed}},
  \bibinfo {author} {\bibfnamefont {L.}~\bibnamefont {Sun}}, \bibinfo {author}
  {\bibfnamefont {B.~R.}\ \bibnamefont {Johnson}}, \bibinfo {author}
  {\bibfnamefont {J.~M.}\ \bibnamefont {Chow}}, \bibinfo {author}
  {\bibfnamefont {J.~M.}\ \bibnamefont {Gambetta}}, \bibinfo {author}
  {\bibfnamefont {L.}~\bibnamefont {Frunzio}}, \bibinfo {author} {\bibfnamefont
  {S.~M.}\ \bibnamefont {Girvin}}, \bibinfo {author} {\bibfnamefont {M.~H.}\
  \bibnamefont {Devoret}}, \ and\ \bibinfo {author} {\bibfnamefont {R.~J.}\
  \bibnamefont {Schoelkopf}},\ }\Doi {10.1038/nature09416} {\bibfield
  {journal} {\bibinfo  {journal} {Nature},\ }\textbf {\bibinfo {volume}
  {467}},\ \bibinfo {pages} {574} (\bibinfo {year} {2010})},\ ISSN \bibinfo
  {issn} {1476-4687}\BibitemShut {NoStop}%
\bibitem [{\citenamefont {Collini}\ \emph {et~al.}(2010)\citenamefont
  {Collini}, \citenamefont {Wong}, \citenamefont {Wilk}, \citenamefont {Curmi},
  \citenamefont {Brumer},\ and\ \citenamefont {Scholes}}]{Collini2010}%
  \BibitemOpen
  \bibfield  {author} {\bibinfo {author} {\bibfnamefont {E.}~\bibnamefont
  {Collini}}, \bibinfo {author} {\bibfnamefont {C.~Y.}\ \bibnamefont {Wong}},
  \bibinfo {author} {\bibfnamefont {K.~E.}\ \bibnamefont {Wilk}}, \bibinfo
  {author} {\bibfnamefont {P.~M.~G.}\ \bibnamefont {Curmi}}, \bibinfo {author}
  {\bibfnamefont {P.}~\bibnamefont {Brumer}}, \ and\ \bibinfo {author}
  {\bibfnamefont {G.~D.}\ \bibnamefont {Scholes}},\ }\Doi {10.1038/nature08811}
  {\bibfield  {journal} {\bibinfo  {journal} {Nature},\ }\textbf {\bibinfo
  {volume} {463}},\ \bibinfo {pages} {644} (\bibinfo {year} {2010})},\ ISSN
  \bibinfo {issn} {1476-4687}\BibitemShut {NoStop}%
\bibitem [{\citenamefont {Semi\~{a}o}\ \emph {et~al.}(2010)\citenamefont
  {Semi\~{a}o}, \citenamefont {Furuya},\ and\ \citenamefont
  {Milburn}}]{Semiao2010}%
  \BibitemOpen
  \bibfield  {author} {\bibinfo {author} {\bibfnamefont {F.}~\bibnamefont
  {Semi\~{a}o}}, \bibinfo {author} {\bibfnamefont {K.}~\bibnamefont {Furuya}},
  \ and\ \bibinfo {author} {\bibfnamefont {G.}~\bibnamefont {Milburn}},\ }\href
  {http://iopscience.iop.org/1367-2630/12/8/083033} {\bibfield  {journal}
  {\bibinfo  {journal} {New J. Phys.},\ }\textbf {\bibinfo {volume} {12}},\
  \bibinfo {pages} {083033} (\bibinfo {year} {2010})}\BibitemShut {NoStop}%
\bibitem [{\citenamefont {Chin}\ \emph {et~al.}(2013)\citenamefont {Chin},
  \citenamefont {Prior}, \citenamefont {Rosenbach}, \citenamefont
  {Caycedo-Soler}, \citenamefont {Huelga},\ and\ \citenamefont
  {Plenio}}]{Chin}%
  \BibitemOpen
  \bibfield  {author} {\bibinfo {author} {\bibfnamefont {A.~W.}\ \bibnamefont
  {Chin}}, \bibinfo {author} {\bibfnamefont {J.}~\bibnamefont {Prior}},
  \bibinfo {author} {\bibfnamefont {R.}~\bibnamefont {Rosenbach}}, \bibinfo
  {author} {\bibfnamefont {F.}~\bibnamefont {Caycedo-Soler}}, \bibinfo {author}
  {\bibfnamefont {S.~F.}\ \bibnamefont {Huelga}}, \ and\ \bibinfo {author}
  {\bibfnamefont {M.~B.}\ \bibnamefont {Plenio}},\ }\href
  {http://arxiv.org/abs/1203.0776
  http://www.nature.com/doifinder/10.1038/nphys2515} {\bibfield  {journal}
  {\bibinfo  {journal} {Nat. Phys.}} (\bibinfo {year} {2013})},\ ISSN \bibinfo
  {issn} {1745-2473}\BibitemShut {NoStop}%
\bibitem [{\citenamefont {Carmichael}\ and\ \citenamefont
  {Walls}(1973)}]{Carmichael1973}%
  \BibitemOpen
  \bibfield  {author} {\bibinfo {author} {\bibfnamefont {H.~J.}\ \bibnamefont
  {Carmichael}}\ and\ \bibinfo {author} {\bibfnamefont {D.~F.}\ \bibnamefont
  {Walls}},\ }\Doi {10.1088/0305-4470/6/10/014} {\bibfield  {journal} {\bibinfo
   {journal} {J. Phys. A.: Math. Gen},\ }\textbf {\bibinfo {volume} {6}},\
  \bibinfo {pages} {1552} (\bibinfo {year} {1973})},\ ISSN \bibinfo {issn}
  {0301-0015}\BibitemShut {NoStop}%
\bibitem [{\citenamefont {Cresser}(1992)}]{Cresser1992}%
  \BibitemOpen
  \bibfield  {author} {\bibinfo {author} {\bibfnamefont {J.~D.}\ \bibnamefont
  {Cresser}},\ }\Doi {10.1080/09500349214552211} {\bibfield  {journal}
  {\bibinfo  {journal} {J. Mod. Opt.},\ }\textbf {\bibinfo {volume} {39}},\
  \bibinfo {pages} {2187} (\bibinfo {year} {1992})},\ ISSN \bibinfo {issn}
  {0950-0340}\BibitemShut {NoStop}%
\bibitem [{\citenamefont {Wilson-Rae}\ and\ \citenamefont
  {Imamoglu}(2002)}]{Wilson-Rae2002}%
  \BibitemOpen
  \bibfield  {author} {\bibinfo {author} {\bibfnamefont {I.}~\bibnamefont
  {Wilson-Rae}}\ and\ \bibinfo {author} {\bibfnamefont {A.}~\bibnamefont
  {Imamoglu}},\ }\Doi {10.1103/PhysRevB.65.235311} {\bibfield  {journal}
  {\bibinfo  {journal} {Phys. Rev. B},\ }\textbf {\bibinfo {volume} {65}},\
  \bibinfo {pages} {235311} (\bibinfo {year} {2002})},\ ISSN \bibinfo {issn}
  {0163-1829}\BibitemShut {NoStop}%
\bibitem [{\citenamefont {F\"{o}rstner}\ \emph {et~al.}(2003)\citenamefont
  {F\"{o}rstner}, \citenamefont {Weber}, \citenamefont {Danckwerts},\ and\
  \citenamefont {Knorr}}]{Forstner2003}%
  \BibitemOpen
  \bibfield  {author} {\bibinfo {author} {\bibfnamefont {J.}~\bibnamefont
  {F\"{o}rstner}}, \bibinfo {author} {\bibfnamefont {C.}~\bibnamefont {Weber}},
  \bibinfo {author} {\bibfnamefont {J.}~\bibnamefont {Danckwerts}}, \ and\
  \bibinfo {author} {\bibfnamefont {A.}~\bibnamefont {Knorr}},\ }\Doi
  {10.1103/PhysRevLett.91.127401} {\bibfield  {journal} {\bibinfo  {journal}
  {Phys. Rev. Lett.},\ }\textbf {\bibinfo {volume} {91}},\ \bibinfo {pages}
  {127401} (\bibinfo {year} {2003})},\ ISSN \bibinfo {issn}
  {0031-9007}\BibitemShut {NoStop}%
\bibitem [{\citenamefont {Vagov}\ \emph {et~al.}(2007)\citenamefont {Vagov},
  \citenamefont {Croitoru}, \citenamefont {Axt}, \citenamefont {Kuhn},\ and\
  \citenamefont {Peeters}}]{Vagov2007a}%
  \BibitemOpen
  \bibfield  {author} {\bibinfo {author} {\bibfnamefont {A.}~\bibnamefont
  {Vagov}}, \bibinfo {author} {\bibfnamefont {M.}~\bibnamefont {Croitoru}},
  \bibinfo {author} {\bibfnamefont {V.}~\bibnamefont {Axt}}, \bibinfo {author}
  {\bibfnamefont {T.}~\bibnamefont {Kuhn}}, \ and\ \bibinfo {author}
  {\bibfnamefont {F.}~\bibnamefont {Peeters}},\ }\Doi
  {10.1103/PhysRevLett.98.227403} {\bibfield  {journal} {\bibinfo  {journal}
  {Phys. Rev. Lett.},\ }\textbf {\bibinfo {volume} {98}},\ \bibinfo {pages}
  {227403} (\bibinfo {year} {2007})},\ ISSN \bibinfo {issn}
  {0031-9007}\BibitemShut {NoStop}%
\bibitem [{\citenamefont {Ramsay}\ \emph
  {et~al.}(2010){\natexlab{a}}\citenamefont {Ramsay}, \citenamefont {Gopal},
  \citenamefont {Gauger}, \citenamefont {Nazir}, \citenamefont {Lovett},
  \citenamefont {Fox},\ and\ \citenamefont {Skolnick}}]{Ramsay2010a}%
  \BibitemOpen
  \bibfield  {author} {\bibinfo {author} {\bibfnamefont {A.~J.}\ \bibnamefont
  {Ramsay}}, \bibinfo {author} {\bibfnamefont {A.~V.}\ \bibnamefont {Gopal}},
  \bibinfo {author} {\bibfnamefont {E.~M.}\ \bibnamefont {Gauger}}, \bibinfo
  {author} {\bibfnamefont {A.}~\bibnamefont {Nazir}}, \bibinfo {author}
  {\bibfnamefont {B.~W.}\ \bibnamefont {Lovett}}, \bibinfo {author}
  {\bibfnamefont {A.~M.}\ \bibnamefont {Fox}}, \ and\ \bibinfo {author}
  {\bibfnamefont {M.~S.}\ \bibnamefont {Skolnick}},\ }\Doi
  {10.1103/PhysRevLett.104.017402} {\bibfield  {journal} {\bibinfo  {journal}
  {Phys. Rev. Lett.},\ }\textbf {\bibinfo {volume} {104}},\ \bibinfo {pages}
  {017402} (\bibinfo {year} {2010}{\natexlab{a}})},\ ISSN \bibinfo {issn}
  {0031-9007}\BibitemShut {NoStop}%
\bibitem [{\citenamefont {Ramsay}\ \emph
  {et~al.}(2010){\natexlab{b}}\citenamefont {Ramsay}, \citenamefont {Godden},
  \citenamefont {Boyle}, \citenamefont {Gauger}, \citenamefont {Nazir},
  \citenamefont {Lovett}, \citenamefont {Fox},\ and\ \citenamefont
  {Skolnick}}]{Ramsay2010}%
  \BibitemOpen
  \bibfield  {author} {\bibinfo {author} {\bibfnamefont {A.}~\bibnamefont
  {Ramsay}}, \bibinfo {author} {\bibfnamefont {T.~M.}\ \bibnamefont {Godden}},
  \bibinfo {author} {\bibfnamefont {S.~J.}\ \bibnamefont {Boyle}}, \bibinfo
  {author} {\bibfnamefont {E.~M.}\ \bibnamefont {Gauger}}, \bibinfo {author}
  {\bibfnamefont {A.}~\bibnamefont {Nazir}}, \bibinfo {author} {\bibfnamefont
  {B.~W.}\ \bibnamefont {Lovett}}, \bibinfo {author} {\bibfnamefont {A.~M.}\
  \bibnamefont {Fox}}, \ and\ \bibinfo {author} {\bibfnamefont {M.~S.}\
  \bibnamefont {Skolnick}},\ }\Doi {10.1103/PhysRevLett.105.177402} {\bibfield
  {journal} {\bibinfo  {journal} {Phys. Rev. Lett.},\ }\textbf {\bibinfo
  {volume} {105}},\ \bibinfo {pages} {177402} (\bibinfo {year}
  {2010}{\natexlab{b}})},\ ISSN \bibinfo {issn} {0031-9007}\BibitemShut
  {NoStop}%
\bibitem [{\citenamefont {McCutcheon}\ and\ \citenamefont
  {Nazir}(2010)}]{McCutcheon2010a}%
  \BibitemOpen
  \bibfield  {author} {\bibinfo {author} {\bibfnamefont {D.~P.~S.}\
  \bibnamefont {McCutcheon}}\ and\ \bibinfo {author} {\bibfnamefont
  {A.}~\bibnamefont {Nazir}},\ }\Doi {10.1088/1367-2630/12/11/113042}
  {\bibfield  {journal} {\bibinfo  {journal} {New J. Phys.},\ }\textbf
  {\bibinfo {volume} {12}},\ \bibinfo {pages} {113042} (\bibinfo {year}
  {2010})},\ ISSN \bibinfo {issn} {1367-2630}\BibitemShut {NoStop}%
\bibitem [{\citenamefont {Roy}\ and\ \citenamefont {Hughes}(2011)}]{Roy2011}%
  \BibitemOpen
  \bibfield  {author} {\bibinfo {author} {\bibfnamefont {C.}~\bibnamefont
  {Roy}}\ and\ \bibinfo {author} {\bibfnamefont {S.}~\bibnamefont {Hughes}},\
  }\Doi {10.1103/PhysRevLett.106.247403} {\bibfield  {journal} {\bibinfo
  {journal} {Phys. Rev. Lett.},\ }\textbf {\bibinfo {volume} {106}},\ \bibinfo
  {pages} {247403} (\bibinfo {year} {2011})},\ ISSN \bibinfo {issn}
  {0031-9007}\BibitemShut {NoStop}%
\bibitem [{\citenamefont {L\"{u}ker}\ \emph {et~al.}(2012)\citenamefont
  {L\"{u}ker}, \citenamefont {Gawarecki}, \citenamefont {Reiter}, \citenamefont
  {Grodecka-Grad}, \citenamefont {Axt}, \citenamefont {Machnikowski},\ and\
  \citenamefont {Kuhn}}]{Luker2012}%
  \BibitemOpen
  \bibfield  {author} {\bibinfo {author} {\bibfnamefont {S.}~\bibnamefont
  {L\"{u}ker}}, \bibinfo {author} {\bibfnamefont {K.}~\bibnamefont
  {Gawarecki}}, \bibinfo {author} {\bibfnamefont {D.~E.}\ \bibnamefont
  {Reiter}}, \bibinfo {author} {\bibfnamefont {A.}~\bibnamefont
  {Grodecka-Grad}}, \bibinfo {author} {\bibfnamefont {V.~M.}\ \bibnamefont
  {Axt}}, \bibinfo {author} {\bibfnamefont {P.}~\bibnamefont {Machnikowski}}, \
  and\ \bibinfo {author} {\bibfnamefont {T.}~\bibnamefont {Kuhn}},\ }\Doi
  {10.1103/PhysRevB.85.121302} {\bibfield  {journal} {\bibinfo  {journal}
  {Phys. Rev. B},\ }\textbf {\bibinfo {volume} {85}},\ \bibinfo {pages}
  {121302} (\bibinfo {year} {2012})},\ ISSN \bibinfo {issn}
  {1098-0121}\BibitemShut {NoStop}%
\bibitem [{\citenamefont {Eastham}\ \emph {et~al.}()\citenamefont {Eastham},
  \citenamefont {Spracklen},\ and\ \citenamefont {Keeling}}]{Eastham2012}%
  \BibitemOpen
  \bibfield  {author} {\bibinfo {author} {\bibfnamefont {P.~R.}\ \bibnamefont
  {Eastham}}, \bibinfo {author} {\bibfnamefont {A.~O.}\ \bibnamefont
  {Spracklen}}, \ and\ \bibinfo {author} {\bibfnamefont {J.}~\bibnamefont
  {Keeling}},\ }\href@noop {} {}\Eprint {http://arxiv.org/abs/1208.5001}
  {arXiv:1208.5001} \BibitemShut {NoStop}%
\bibitem [{\citenamefont {Reiter}\ \emph {et~al.}(2012)\citenamefont {Reiter},
  \citenamefont {L\"{u}ker}, \citenamefont {Gawarecki}, \citenamefont
  {Grodecka-Grad}, \citenamefont {Machnikowski}, \citenamefont {Axt},\ and\
  \citenamefont {Kuhn}}]{Reiter2012}%
  \BibitemOpen
  \bibfield  {author} {\bibinfo {author} {\bibfnamefont {D.~E.}\ \bibnamefont
  {Reiter}}, \bibinfo {author} {\bibfnamefont {S.}~\bibnamefont {L\"{u}ker}},
  \bibinfo {author} {\bibfnamefont {K.}~\bibnamefont {Gawarecki}}, \bibinfo
  {author} {\bibfnamefont {A.}~\bibnamefont {Grodecka-Grad}}, \bibinfo {author}
  {\bibfnamefont {P.}~\bibnamefont {Machnikowski}}, \bibinfo {author}
  {\bibfnamefont {V.~M.}\ \bibnamefont {Axt}}, \ and\ \bibinfo {author}
  {\bibfnamefont {T.}~\bibnamefont {Kuhn}},\ }\href
  {http://arxiv.org/abs/1207.6660} {\bibfield  {journal} {\bibinfo  {journal}
  {Acta. Phys. Pol. A},\ }\textbf {\bibinfo {volume} {122}},\ \bibinfo {pages}
  {1065} (\bibinfo {year} {2012})}\BibitemShut {NoStop}%
\bibitem [{\citenamefont {Nataf}\ and\ \citenamefont
  {Ciuti}(2010)}]{Nataf:Nogo}%
  \BibitemOpen
  \bibfield  {author} {\bibinfo {author} {\bibfnamefont {P.}~\bibnamefont
  {Nataf}}\ and\ \bibinfo {author} {\bibfnamefont {C.}~\bibnamefont {Ciuti}},\
  }\Doi {10.1038/ncomms1069} {\bibfield  {journal} {\bibinfo  {journal} {Nat.
  Commun.},\ }\textbf {\bibinfo {volume} {1}},\ \bibinfo {pages} {72} (\bibinfo
  {year} {2010})},\ ISSN \bibinfo {issn} {2041-1723}\BibitemShut {NoStop}%
\bibitem [{Note1()}]{Note1}%
  \BibitemOpen
  \bibinfo {note} {There is some debate regarding whether the transition can be
  achieved for circuit QED~\cite
  {Rzazewski1975a,*Viehmann:SR,*Vukics2012}.}\BibitemShut {Stop}%
\bibitem [{\citenamefont {Ciuti}\ and\ \citenamefont
  {Carusotto}(2006)}]{Ciuti2006}%
  \BibitemOpen
  \bibfield  {author} {\bibinfo {author} {\bibfnamefont {C.}~\bibnamefont
  {Ciuti}}\ and\ \bibinfo {author} {\bibfnamefont {I.}~\bibnamefont
  {Carusotto}},\ }\Doi {10.1103/PhysRevA.74.033811} {\bibfield  {journal}
  {\bibinfo  {journal} {Phys. Rev. A},\ }\textbf {\bibinfo {volume} {74}},\
  \bibinfo {pages} {033811} (\bibinfo {year} {2006})},\ ISSN \bibinfo {issn}
  {1050-2947}\BibitemShut {NoStop}%
\bibitem [{\citenamefont {Nigg}\ \emph {et~al.}(2012)\citenamefont {Nigg},
  \citenamefont {Paik}, \citenamefont {Vlastakis}, \citenamefont {Kirchmair},
  \citenamefont {Shankar}, \citenamefont {Frunzio}, \citenamefont {Devoret},
  \citenamefont {Schoelkopf},\ and\ \citenamefont
  {Girvin}}]{PhysRevLett.108.240502}%
  \BibitemOpen
  \bibfield  {author} {\bibinfo {author} {\bibfnamefont {S.~E.}\ \bibnamefont
  {Nigg}}, \bibinfo {author} {\bibfnamefont {H.}~\bibnamefont {Paik}}, \bibinfo
  {author} {\bibfnamefont {B.}~\bibnamefont {Vlastakis}}, \bibinfo {author}
  {\bibfnamefont {G.}~\bibnamefont {Kirchmair}}, \bibinfo {author}
  {\bibfnamefont {S.}~\bibnamefont {Shankar}}, \bibinfo {author} {\bibfnamefont
  {L.}~\bibnamefont {Frunzio}}, \bibinfo {author} {\bibfnamefont {M.~H.}\
  \bibnamefont {Devoret}}, \bibinfo {author} {\bibfnamefont {R.~J.}\
  \bibnamefont {Schoelkopf}}, \ and\ \bibinfo {author} {\bibfnamefont {S.~M.}\
  \bibnamefont {Girvin}},\ }\Doi {10.1103/PhysRevLett.108.240502} {\bibfield
  {journal} {\bibinfo  {journal} {Phys. Rev. Lett.},\ }\textbf {\bibinfo
  {volume} {108}},\ \bibinfo {pages} {240502} (\bibinfo {year}
  {2012})}\BibitemShut {NoStop}%
\bibitem [{\citenamefont {Mlynek}\ \emph {et~al.}(2012)\citenamefont {Mlynek},
  \citenamefont {Abdumalikov}, \citenamefont {Fink}, \citenamefont {Steffen},
  \citenamefont {Baur}, \citenamefont {Lang}, \citenamefont {van Loo},\ and\
  \citenamefont {Wallraff}}]{Mlynek2012}%
  \BibitemOpen
  \bibfield  {author} {\bibinfo {author} {\bibfnamefont {J.}~\bibnamefont
  {Mlynek}}, \bibinfo {author} {\bibfnamefont {A.}~\bibnamefont {Abdumalikov}},
  \bibinfo {author} {\bibfnamefont {J.}~\bibnamefont {Fink}}, \bibinfo {author}
  {\bibfnamefont {L.}~\bibnamefont {Steffen}}, \bibinfo {author} {\bibfnamefont
  {M.}~\bibnamefont {Baur}}, \bibinfo {author} {\bibfnamefont {C.}~\bibnamefont
  {Lang}}, \bibinfo {author} {\bibfnamefont {A.~F.}\ \bibnamefont {van Loo}}, \
  and\ \bibinfo {author} {\bibfnamefont {A.}~\bibnamefont {Wallraff}},\ }\Doi
  {10.1103/PhysRevA.86.053838} {\bibfield  {journal} {\bibinfo  {journal}
  {Phys. Rev. A},\ }\textbf {\bibinfo {volume} {86}},\ \bibinfo {pages}
  {053838} (\bibinfo {year} {2012})},\ ISSN \bibinfo {issn}
  {1050-2947}\BibitemShut {NoStop}%
\bibitem [{\citenamefont {Yoshihara}\ \emph {et~al.}(2006)\citenamefont
  {Yoshihara}, \citenamefont {Harrabi}, \citenamefont {Niskanen}, \citenamefont
  {Nakamura},\ and\ \citenamefont {Tsai}}]{Yoshihara2006a}%
  \BibitemOpen
  \bibfield  {author} {\bibinfo {author} {\bibfnamefont {F.}~\bibnamefont
  {Yoshihara}}, \bibinfo {author} {\bibfnamefont {K.}~\bibnamefont {Harrabi}},
  \bibinfo {author} {\bibfnamefont {A.}~\bibnamefont {Niskanen}}, \bibinfo
  {author} {\bibfnamefont {Y.}~\bibnamefont {Nakamura}}, \ and\ \bibinfo
  {author} {\bibfnamefont {J.}~\bibnamefont {Tsai}},\ }\Doi
  {10.1103/PhysRevLett.97.167001} {\bibfield  {journal} {\bibinfo  {journal}
  {Phys. Rev. Lett.},\ }\textbf {\bibinfo {volume} {97}},\ \bibinfo {pages}
  {167001} (\bibinfo {year} {2006})},\ ISSN \bibinfo {issn}
  {0031-9007}\BibitemShut {NoStop}%
\bibitem [{\citenamefont {Breuer}\ and\ \citenamefont
  {Petruccione}(2007)}]{Breuer2007}%
  \BibitemOpen
  \bibfield  {author} {\bibinfo {author} {\bibfnamefont {H.-P.}\ \bibnamefont
  {Breuer}}\ and\ \bibinfo {author} {\bibfnamefont {F.}~\bibnamefont
  {Petruccione}},\ }\href@noop {} {\emph {\bibinfo {title} {{The Theory of Open
  Quantum Systems}}}}\ (\bibinfo  {publisher} {Oxford University Press},\
  \bibinfo {address} {Oxford},\ \bibinfo {year} {2007})\ ISBN \bibinfo {isbn}
  {0199213909}\BibitemShut {NoStop}%
\bibitem [{\citenamefont {Gawarecki}\ \emph {et~al.}(2012)\citenamefont
  {Gawarecki}, \citenamefont {L\"{u}ker}, \citenamefont {Reiter}, \citenamefont
  {Kuhn}, \citenamefont {Gl\"{a}ssl}, \citenamefont {Axt}, \citenamefont
  {Grodecka-Grad},\ and\ \citenamefont {Machnikowski}}]{Gawarecki2012}%
  \BibitemOpen
  \bibfield  {author} {\bibinfo {author} {\bibfnamefont {K.}~\bibnamefont
  {Gawarecki}}, \bibinfo {author} {\bibfnamefont {S.}~\bibnamefont
  {L\"{u}ker}}, \bibinfo {author} {\bibfnamefont {D.}~\bibnamefont {Reiter}},
  \bibinfo {author} {\bibfnamefont {T.}~\bibnamefont {Kuhn}}, \bibinfo {author}
  {\bibfnamefont {M.}~\bibnamefont {Gl\"{a}ssl}}, \bibinfo {author}
  {\bibfnamefont {V.}~\bibnamefont {Axt}}, \bibinfo {author} {\bibfnamefont
  {A.}~\bibnamefont {Grodecka-Grad}}, \ and\ \bibinfo {author} {\bibfnamefont
  {P.}~\bibnamefont {Machnikowski}},\ }\Doi {10.1103/PhysRevB.86.235301}
  {\bibfield  {journal} {\bibinfo  {journal} {Phys. Rev. B},\ }\textbf
  {\bibinfo {volume} {86}},\ \bibinfo {pages} {235301} (\bibinfo {year}
  {2012})},\ ISSN \bibinfo {issn} {1098-0121}\BibitemShut {NoStop}%
\bibitem [{\citenamefont {Dumcke}\ and\ \citenamefont
  {Spohn}(1979)}]{Dmcke1979}%
  \BibitemOpen
  \bibfield  {author} {\bibinfo {author} {\bibfnamefont {R.}~\bibnamefont
  {Dumcke}}\ and\ \bibinfo {author} {\bibfnamefont {H.}~\bibnamefont {Spohn}},\
  }\Doi {10.1007/BF01325208} {\bibfield  {journal} {\bibinfo  {journal} {Z.
  Phys. B},\ }\textbf {\bibinfo {volume} {34}},\ \bibinfo {pages} {419}
  (\bibinfo {year} {1979})},\ ISSN \bibinfo {issn} {0340-224X}\BibitemShut
  {NoStop}%
\bibitem [{\citenamefont {Rzazewski}\ \emph {et~al.}(1975)\citenamefont
  {Rzazewski}, \citenamefont {W\'{o}dkiewicz},\ and\ \citenamefont
  {Zakowicz}}]{Rzazewski1975a}%
  \BibitemOpen
  \bibfield  {author} {\bibinfo {author} {\bibfnamefont {K.}~\bibnamefont
  {Rzazewski}}, \bibinfo {author} {\bibfnamefont {K.}~\bibnamefont
  {W\'{o}dkiewicz}}, \ and\ \bibinfo {author} {\bibfnamefont {W.}~\bibnamefont
  {Zakowicz}},\ }\Doi {10.1103/PhysRevLett.35.432} {\bibfield  {journal}
  {\bibinfo  {journal} {Phys. Rev. Lett.},\ }\textbf {\bibinfo {volume} {35}},\
  \bibinfo {pages} {432} (\bibinfo {year} {1975})},\ ISSN \bibinfo {issn}
  {0031-9007}\BibitemShut {NoStop}%
\bibitem [{\citenamefont {Viehmann}\ \emph {et~al.}(2011)\citenamefont
  {Viehmann}, \citenamefont {von Delft},\ and\ \citenamefont
  {Marquardt}}]{Viehmann:SR}%
  \BibitemOpen
  \bibfield  {author} {\bibinfo {author} {\bibfnamefont {O.}~\bibnamefont
  {Viehmann}}, \bibinfo {author} {\bibfnamefont {J.}~\bibnamefont {von Delft}},
  \ and\ \bibinfo {author} {\bibfnamefont {F.}~\bibnamefont {Marquardt}},\
  }\Doi {10.1103/PhysRevLett.107.113602} {\bibfield  {journal} {\bibinfo
  {journal} {Phys. Rev. Lett.},\ }\textbf {\bibinfo {volume} {107}},\ \bibinfo
  {pages} {113602} (\bibinfo {year} {2011})},\ ISSN \bibinfo {issn}
  {0031-9007}\BibitemShut {NoStop}%
\bibitem [{\citenamefont {Vukics}\ and\ \citenamefont
  {Domokos}(2012)}]{Vukics2012}%
  \BibitemOpen
  \bibfield  {author} {\bibinfo {author} {\bibfnamefont {A.}~\bibnamefont
  {Vukics}}\ and\ \bibinfo {author} {\bibfnamefont {P.}~\bibnamefont
  {Domokos}},\ }\href {http://arxiv.org/abs/1206.0752
  http://link.aps.org/doi/10.1103/PhysRevA.86.053807} {\bibfield  {journal}
  {\bibinfo  {journal} {Phys. Rev. A},\ }\textbf {\bibinfo {volume} {86}},\
  \bibinfo {pages} {053807} (\bibinfo {year} {2012})},\ ISSN \bibinfo {issn}
  {1050-2947}\BibitemShut {NoStop}%
\end{thebibliography}
%

\end{document}